\documentclass[fleqn,twoside]{article}%
\usepackage{amsmath}
\usepackage{amsfonts}
\usepackage{amssymb}
\usepackage{graphicx}%
\def\be{\begin{equation}}

\def\ee{\end{equation}}

\def\bi{\bibitem}

\begin{document}

\title{Viability of Noether symmetry of $F(R)$ theory of gravity}

\author{Kaushik Sarkar$^{\ast}$, Nayem Sk.$^\dag$, Subhra Debnath$^{\ddag,\star}$,  and Abhik Kumar Sanyal$^{\ddag,\S}$}
\maketitle

\noindent

\begin{center}
$\ast$Dept. of Physics, University of Kalyani, Nadia, India - 741235\\
$\dag$Dept. of Physics, Ranitala High School, Murshidabad, India - 742135\\
$\ddag$Dept. of Physics, Jangipur College, Murshidabad, India - 742213\\

\end{center}

\footnotetext[1]{
Electronic address:\\
\noindent $^{\ast}$sarkarkaushik.rng@gmail.com\\
\noindent $\dag$nayemsk1981@gmail.com\\
\noindent $^{\star}$subhra\_ dbnth@yahoo.com\\
\noindent $^{\S}$sanyal\_ ak@yahoo.com\\}

\begin{abstract}

\noindent Canonization of $F(R)$ theory of gravity to explore Noether symmetry is performed treating $R - 6(\frac{\ddot a}{a} + \frac{\dot a^2}{a^2} + \frac{k}{a^2}) = 0$ as a constraint of the theory in Robertson-Walker space-time, which implies that $R$ is taken as an auxiliary variable. Although it yields correct field equations, Noether symmetry does not allow linear term in the action, and as such does not produce a viable cosmological model. Here, we show that this technique of exploring  Noether symmetry does not allow even a non-linear form of $F(R)$, if the configuration space is enlarged by including a scalar field in addition, or taking anisotropic models into account. Surprisingly enough, it does not reproduce the symmetry that already exists in the literature \cite{18} for scalar tensor theory of gravity in the presence of $R^2$ term. Thus, $R$ can not be treated as an auxiliary variable and hence Noether symmetry of arbitrary form of $F(R)$ theory of gravity remains obscure. However, there exists in general, a conserved current for $F(R)$ theory of gravity in the presence of a non-minimally coupled scalar-tensor theory \cite{27,28}. Here, we briefly expatiate the non-Noether conserved current and cite an example to reveal its importance in finding cosmological solution for such an action, taking $F(R) \propto R^{\frac{3}{2}}$.
\end{abstract}

\section{Introduction}

Increasing interest in $F(R)$ theory of gravity (see \cite{1} for a recent review) initiates to probe deeply into it and to explore its merits-demerits from different angles. Cosmological consequences of $F(R)$ theory of gravity have been explored \cite{2} taking into account different sets of field equations which appear through the metric variation and Palatini variation formalisms. Although, both set of field equations lead to late time cosmic acceleration without the requirement of dark energy in the form of scalar fields, different results have emerged following these two different techniques. For example, unification of early inflation and late time acceleration \cite{3} on one hand, and violent instability \cite{7}, cosmologically non-viability \cite{8}, big bang nucleosynthesis and fifth-force constraints altogether \cite{9}, on the other, are the outcome of metric formalism. On the contrary, no instability appears as such in Palatini formalism \cite{10} and it leads to correct Newtonian limit \cite{11}, while curvature corrections are found to induce effective pressure gradients which create problem in the formation of large-scale structure \cite{12}. Further, even density perturbation in the two techniques, produce different results \cite{13}. Nevertheless, it has been shown that under conformal transformation the two formalisms are the same dynamically \cite{14} and also are equivalent for a large class of theories \cite{15}. Despite such contrasting results on one hand and equivalence of the two formalisms on the other, at the end it is required to choose some particular form of $F(R)$ out of indefinitely many. This is only possible by imposing Noether symmetry. In the present work we discuss this issue in the metric variation formalism.\\

\noindent
Noether symmetry when applied for the first time in scalar-tensor theory of gravity \cite{16}, the idea was to find a form of the potential that might give rise to a cyclic co-ordinate and hence a conserved current. That is, out of indefinitely many choice of the potential, Noether symmetry selects one, corresponding to which there exists a cyclic coordinate that can't be found a-priori through inspection. In the Robertson-Walker background metric the exponential form of the potential thus found was rather encouraging, because it could trigger inflation in the early Universe. Since then, it has been applied in nonminimally coupled scalar-tensor theory of gravity both in the background of isotropic and several anisotropic models \cite{17}, together with nonminimally coupled scalar-tensor theory in the presence of $R^2$ term in the action \cite{18} and recently, to classify different dark energy models \cite{19}.\\

\noindent
To explore Noether symmetry, it is first required to express the action in canonical form, which is possible by introducing auxiliary variable in higher order theory of gravity. This has been attempted by several authors \cite{21} for $F(R)$ theory of gravity, treating $R - 6(\frac{\ddot a}{a} + \frac{\dot a^2}{a^2} + \frac{k}{a^2}) = 0$ as a constraint in the Robertson-Walker minisuperspace model. In the process, the Lagrangian is spanned by a set of configuration space variables $(a, \dot a, R, \dot R)$, i.e., $R$ is treated as an auxiliary variable and the result is $F(R) \propto R^{\frac{3}{2}}$ in the vacuum and matter dominated era. Such a form of $F(R)$ produces $a \propto t^{\frac{3}{4}}$ instead of the standard $\sqrt{t}$ in the radiation dominated era \cite{22}. This creates problem in explaining BBN, since Universe expands at a much faster rate than the standard Friedmann model. Further, it has been shown that \cite{22} under appropriate choice of variable $h_{ij} = a^2 = z$, required for canonical quantization of higher order theory of gravity in the Robertson-Walker metric \cite{22a}, $z$ becomes cyclic for $F(R) =  R^{\frac{3}{2}}$. It means $R^{\frac{3}{2}}$ has inbuilt Noether symmetry for $F(R)$ theory in vacuum or in matter dominated era. Now, scalar field appears naturally along with higher order curvature invariant terms in the weak energy limit of string theories \cite{23}, $4$-dimensional Brane world effective action \cite{24}, supergravity theories \cite{25} and in the Chern-Simons gravitational theories \cite{26}. Therefore, adding more degrees of freedom in the form of a scalar field, $z$ is no longer cyclic for $F(R) =  R^{\frac{3}{2}}$ and situation might improve, yielding a better form of $F(R)$ to explain presently available cosmological data. Additionally, the symmetry already obtained for $F(R) = f(\phi)R + \beta R^2$, in the presence of a scalar field \cite{18} must be reproduced, if canonization treating $R$ as an auxiliary variable is correct. \\

\noindent
Higher order theory of gravity may be expressed in canonical form choosing other auxiliary variables too, different from $R$. All lead to correct classical field equations but different quantum dynamics. Horowitz \cite{h} suggested to choose auxiliary variable as the derivative of the action with respect to the highest derivative present in the action, i.e., as $Q = \frac{\partial A}{\partial \ddot a}$. This suggestion is flawed, since in that case auxiliary variable may be introduced even in linear theory as shown by Pollock \cite{p} leading to completely wrong quantum dynamics. Later, in a series of works \cite{22a} it was suggested that auxiliary variable should be introduced following the prescription of Horowitz \cite{h} only after removing appropriate total derivative terms from the action. Following this technique, we found Noether symmetry for $F(R) = f(\phi) R + \beta R^2$, $R$ being non-minimally coupled to a scalar field \cite{18}. Thus our intention is to see if same result \cite{18} is reproduced taking $R$ as an auxiliary variable, and if Noether symmetry at all exists for other form of action.\\

\noindent
Thus in the following section 2, we take up a theory of gravity consisting of scalar field being nonminimally coupled with $F(R)$, follow the usual technique of finding Noether symmetry \cite{21} to end up with the result that Noether symmetry for $F_{,RR} \ne 0$ is obscure. In section 3, following the same technique \cite{21}, for a nonminimally coupled scalar-tensor theory of gravity along with $F(R)$ term we obtain the same result that $F_{,RR} \ne 0$ is impossible. This is surprising, since, as already mentioned, we have explored Noether symmetry for the same action, with $F(R) = f(\phi)R + \beta R^2$ and the Brans-Dicke coupling parameter $\omega(\phi) = 1$, in the conformally flat Robertson-Walker metric \cite{18}. In doing so, we have used an auxiliary variable $Q$, different from $R$, and the Lagrangian was spanned by the set of configuration space variables ($a, Q, \phi, \dot a, \dot Q, \dot\phi$), instead. Thus, the absence of Noether symmetry in both the cases automatically raises doubt about the technique of treating $R - 6(\frac{\ddot a}{a} + \frac{\dot a^2}{a^2} + \frac{k}{a^2}) = 0$, as a constraint of the theory. To check if indeed the technique is flawed, in section 4, we take up a set of anisotropic models, which also increases the configuration space variables ($a, b, R, \dot a, \dot b, \dot R$), to explore Noether symmetry of $F(R)$ theory in vacuum. Here again we find that $F(R)$ does not admit any nonlinear form. Thus we conclude that symmetry in isotropic space-time obtained for $F(R) = R^{\frac{3}{2}}$ in vacuum and matter dominated era, was an accident, since it makes $z = a^2$ cyclic, but the technique \cite{21} is flawed, and it is not possible to explore Noether symmetry of $F(R)$ theory of gravity, in general.\\

\noindent
The next question that automatically arises, ``is it possible to find an integral of motion in general for $F(R)$ theory of gravity being coupled with a scalar field, including a linear term in addition?". The answer is yes and already appears in the literature \cite{27}, where a general conserved current in connection with a general higher order theory of gravity coupled with a dilatonic scalar has been explored in a metric independent way. In section 5, we briefly discuss the technique of finding the same and as an example, show how to use such conserved current to find solutions in the context of cosmology. We have given a simple solution of the scale factor directly for the action containing $F(R) = R^{\frac{3}{2}}$ being added to a linear term ($f(\phi)R$) along with a scalar field $\phi$, in the Robertson-Walker metric, both in radiation and matter dominated era. It produces Friedmann type solution in the radiation era and a coasting solution in the matter era, which fits SnIa data perfectly although early deceleration remains absent creating problem in structure formation. Finally, we end up with conclusions.\\

\section{$F(R)$, non-minimally coupled with a scalar field.}
Here, we start with the following action, in which $F(R)$ is non-minimally coupled to a scalar field as,

\be S = \int d^4 x\sqrt{-g}\left[h(\phi) F(R) - \frac{w(\phi)}{2}\phi_{,\mu}\phi^{,\mu} - V(\phi)\right].\ee
In the Robertson-Walker line element,

\be ds^2 = -dt^2+a^2\left[\frac{dr^2}{1-kr^2}+r^2 d\theta^2 + r^2\sin^2\theta d\phi^2\right],\ee

\be R = 6\left(\frac{\ddot a}{a}+\frac{\dot a^2}{a^2}+\frac{k}{a^2}\right).\ee
As already mentioned, the technique \cite{21} under discussion is to treat $R - 6\left(\frac{\ddot a}{a}+\frac{\dot a^2}{a^2}+\frac{k}{a^2}\right) =0$ as a constraint and to introduce it in the action via a Lagrange multiplier $\lambda$ as,

\be S = \int d^4 x\sqrt{-g}\left[h(\phi) F(R) - \lambda\left\{R - 6\left(\frac{\ddot a}{a}+\frac{\dot a^2}{a^2}+\frac{k}{a^2}\right)\right\} - \frac{w(\phi)}{2}\phi_{,\mu}\phi^{,\mu} - V(\phi)\right],\ee
and to vary the action with respect to $R$, which yields,

\be \lambda = h(\phi)F_{,R}(R).\ee
Next step is to substitute the form of $\lambda$ into the above action. After  integration by parts to get rid of the appropriate surface term, the action is automatically expressed in the following non-degenerate canonical form, that can be checked by calculating the determinant of the Hessian ${\mathcal H} = |\Sigma_{i,j} \frac{\partial^2 L}{\partial\dot q_i\partial\dot q_j}|\ne 0$. Thus,

\be S = \int \left[ h a^3(F-R F_{,R})-6 h a\dot a^2F_{,R} -6 h a^2\dot a\dot R F_{,RR}-6 h'a^2\dot a\dot\phi F_{,R}+6k h a F_{,R}+ a^3\left(\frac{w}{2}\dot\phi^2 - V\right)\right]dt.\ee
We now move straight into the process of finding a non-linear form of $F(R)$, along with $w(\phi)$ and $V(\phi)$, by imposing Noether symmetry, $\pounds_{X}L = XL = 0,$ where, $X$ is the vector field and $L$ is the point Lagrangian. So, as usual, let us equate the coefficients of $\dot a^2,\;\dot R^2,\;\dot\phi^2,\;\dot a\dot\phi,\;\dot\phi\dot R,\; \dot R\dot a$ and the terms independent of them, separately to zero to get the following seven equations respectively.

\begin{eqnarray}
&ah(\beta+a\beta_{,a}) F_{,RR} + (\alpha h +2a\alpha_{,a}h+a\gamma h'+a^2\gamma_{,a}h')F_{,R} = 0.\\
&6ha^2F_{,RR}\alpha_{,R}=0.\\
&3(w\alpha-4h'F_{,R}\alpha')+a(w'\gamma+2w\gamma')=0.\\
&a[h'\beta+h\beta']F_{,RR} + [h'(2\alpha+a\alpha_{,a})+2h\alpha'+a(h''\gamma+h'\gamma')]F_{,R}-\frac{1}{6}a^2w\gamma_{,a}=0.\\
&6h\alpha'F_{,RR}+6h'\alpha_{,R}F_{,R} = wa\gamma_{,R}.\\
&\beta h a F_{,RRR}+[2h\alpha+ha\alpha_{,a}+ha\beta_{,R}+h'a\gamma]F_{,RR}+(2h\alpha_{,R}+h'a\gamma_{,R})F_{,R}=0.\\
&\beta ha[6k-a^2R]F_{,RR} + a^2(F-R F_{,R})(3\alpha h + \gamma a h')+6kF_{,R}(\alpha h + \gamma a h')-a^2(3\alpha V + \gamma a V') = 0.
\end{eqnarray}
In the above set of equations and everywhere else dash ($'$) represents derivative with respect to $\phi$. In the following subsection, we try to explore Noether symmetry from the above set of equations (7) through (13).

\subsection {Solutions.}

\noindent
Since, we are only interested in a non-linear form of $F(R)$, i.e., $F_{,RR} \ne 0$, so equation (8) implies $\alpha_{,R} = 0$, i.e., $\alpha = \alpha(a, \phi)$. Hence, equation (11) leads to two distinct cases, which we explore underneath in the subsubsections (2.1.1) and (2.1.2).\\
\subsubsection{Case 1.}
The first case that arises in view of equation (11) is,
\[F_{,RR} = \frac{a w}{6h\alpha'}\gamma_{,R},\]
which can be integrated to yield
\be F_{,R} = \frac{a w}{6h\alpha'}\gamma + m,\;\;\; i.e.,\;\;\;\gamma = \frac{6h\alpha'}{a w}(F_{,R} - m).\ee
where, $m$ is clearly a constant, since $F_{,R}$, is a function of $R$ only. Hence, substituting this form of $\gamma$ and $\gamma'$ in equation (9) we obtain,
\[2hF_{,R}(w'\alpha' - 2w\alpha'') = w^2\alpha - 2m(2wh\alpha''+2wh'\alpha'-w'h\alpha'),\hspace{2.57in}(i)\]
which holds only under the following conditions,
\[2w\alpha'' = w'\alpha' \Longrightarrow \alpha'^2 = \alpha_1(a)^2 w,\hspace{4.24in}(ii)\]
\[w^2\alpha = 2m(2wh\alpha''+2wh'\alpha'-w'h\alpha').\hspace{3.94in}(iii)\]
These pair of equations relate $w$, $\alpha$ and $h$. In fact using $(ii)$, equation $(iii)$ takes the form (Note, $\alpha' = 0 \Rightarrow w = 0$, for which $\phi$ turns out to be a Lagrange multiplier, and so we omit this case),
\[w\alpha = 4mh'\alpha'.\hspace{5.413in}(iv)\]
Finally, equations $(ii)$ and $(iv)$ imply,
\[ \frac{\alpha\alpha'}{\alpha_{1}^2} = 4mh'.\hspace{5.51in}(v)\]
Since, the right hand side of $(v)$ is a function of $\phi$ only, so $\alpha$ must have the separable form, $\alpha = \alpha_1(a)\alpha_2(\phi)$, and thus,
\be \alpha_2^2 = 8 mh\;\;\;\;and\;\;\;w = \alpha_2'^2.\ee
Now using equations (14) and (15), one can express equation (7) as,
\be (\xi + a \xi_{,a})\beta_3 F_{,RR} + \frac{3}{2m}\alpha_{1,a}\alpha_2 {F_{,R}}^2 + \Big[\frac{\alpha_1}{a} + \frac{\alpha_{1,a}}{2}\Big]\alpha_2 F_{,R} = 0.\ee
where, $\beta = \xi(a, \phi)\beta_3(R)$. Equation (16) leads to three possibilities, which we study in the following three sub-cases.\\

\noindent
\textbf{Subcase-1}\\
\[\beta_3 F_{,RR} = n F_{,R},\;\; \alpha \ne \alpha(a),\;\; n a(a\xi)_{,a} + \alpha_{1}\alpha_{2} = 0.\]

\noindent
Substituting the above condition along with the forms of $\gamma$ and $\gamma'$ from equation (14) in Noether equation (10) we arrive at,

\[\left[\frac{6hh'\alpha_1}{\alpha_2'} + \frac{6h'^2\alpha_1}{\alpha_2'}-\frac{6hh'\alpha_1 \alpha_2''}{(\alpha_2')^2}\right]F_{,R}^2\]
\[+ \left[n a (h\xi)' + 2\alpha_1 (\alpha_2 h)' + \alpha_1 \alpha_2'h - m\Big(\frac{6hh'\alpha_1}{\alpha_2'} + \frac{6h'^2\alpha_1}{\alpha_2'}-\frac{6hh'\alpha_1 \alpha_2''}{(\alpha_2')^2}\Big)\right]F_{,R} -\alpha_1\alpha_2'hm = 0\]

\noindent
So, either $\alpha_2' = 0$, or, $m = 0$, yielding $\alpha_2 = 0$. Both the cases imply $\omega = 0$ in view of equation (15), turning out $\phi$ to be a Lagrange multiplier, which is not our concern. Thus we switch over to the following subcase.\\

\noindent
\textbf{Subcase-2}\\
\[ \xi + a \xi_{,a} = 0,\;\;\;\alpha_1 = 0,\]
which is trivial, or,
\[ \xi + a \xi_{,a} = 0,\;\;\;\alpha_2 = 0 \;\;\; and \;\;\; a\alpha_{1,a}+2\alpha_{1} = 0,\]
which is again trivial. So the following subcase is only of importance.\\

\noindent
\textbf{Subcase-3},
\be \beta_3 F_{,RR} = q {F_{,R}}^2,\ee
$q$ being a constant, along with the coefficients,
\be q(\xi + a \xi_{,a}) + \frac{3}{2m}\alpha_{1,a}\alpha_2 = 0\;\;\;and\;\;\;a\alpha_{1,a}+2\alpha_{1} = 0.\ee
i.e., \be \alpha_1 = a^{-2},\;\;\;\;\;\;\alpha_2 = \frac{m q q_2}{3}\beta_2\;\;\;and\;\;\;a^3(\beta_1+a\beta_{1,a})=q_2 \Longrightarrow \beta_1 = -\frac{q_2}{2 a^3}.\ee
In the above we have further separated $\xi(a, \phi)$ as $\xi(a, \phi) = \beta_{1}(a)\beta_{2}(\phi)$. If we now take up equation (10), it leads to,
\[ \left[q a(h \xi)'+ \frac{6\alpha' h h''}{w}+6h'\left(\frac{h\alpha'}{w}\right)'\right]{F_{,R}}^2\]\be+\left[(2\alpha+a\alpha_{,a}+2\alpha')h-\frac{6mh h''\alpha'}{w}-6mh'\left(\frac{h\alpha'}{w}\right)'-a^2 h\left(\frac{\alpha'}{a}\right)_{,a}\right]F_{,R}+ma^2 h\left(\frac{\alpha'}{a}\right)_{,a} = 0.\ee
Since coefficients must vanish separately, so, either $\alpha' = 0 \Longrightarrow w = 0$, which is trivial, as already discussed, or, $\alpha \ne \alpha(a)$. But then, to satisfy equation (18), $\alpha_{1} = 0$, i.e., $\alpha = 0$, which again leads to a trivial situation. Thus, Noether symmetry for $F_{,RR} \ne 0$, remains obscure.\\
\subsubsection{Case 2.}
Equation (11) leads to yet another situation in which, $\alpha' = 0 =\gamma_{,R}$, i.e., $\alpha = \alpha(a)$ and $\gamma = \gamma(a, \phi)$. Thus equation (9) yields,

\be 3\frac{\alpha}{a} + \gamma\frac{w'}{w} +2\gamma'=0.\ee
Now, separating $\beta$ as, $\beta = \xi(a, \phi) \beta_{3}(R)$, equation (7) leads to,

\be \beta_{3} \frac{F_{,RR}}{F_{,R}} = n_1 = -\frac{\alpha+2a\alpha_{,a}+a\frac{h'}{h}(\gamma+a\gamma_{,a})}{a(\xi + a \xi_{,a})},\ee
while equation (12) yields,

\be \beta_{3} \frac{F_{,RRR}}{F_{,RR}}+\beta_{3,R} = n_1 = -\frac{2(\alpha+a\alpha_{,a}+a\frac{h'}{h}\gamma)}{a \xi}.\ee
Left hand side of the above pair of equations are essentially the same, and so we have taken the same separation constant $n_{1}$ for the both. Equation (10) now takes the following form,

\be (h \xi)'\beta_{3}F_{,RR}+\left[(2\frac{\alpha}{a}+\alpha_{,a})h'+2\frac{\alpha'}{a}h+(\gamma h')'\right]F_{,R} - \frac{a}{6}w\gamma_{,a} =0.\ee
At this end, plugging in equation (22) in the above equation, we arrive at,
\be  n_1(h \xi)' +(2\frac{\alpha}{a}+\alpha_{,a})h'+2\frac{\alpha'}{a}h +(\gamma h')' = 0;\;\;\;\gamma=\gamma(\phi).\ee
In this case, since $\gamma$ is only a function of $\phi$, so equation (21) implies
\be\alpha = \alpha(a) = ma \;\;\;and\;\;\;\gamma\frac{w'}{w} + 2\gamma' = -3m.\ee
Thus, equation (23) yields,
\be 2\gamma\frac{h'}{h} + n_1 \xi + 4m = 0,\ee
which clearly indicates that $\xi = \xi(\phi)$. Further, equation (22) leads to,
\be\gamma\frac{h'}{h} + n_1 \xi + 3m = 0.\ee
Last pair of equations (27) and (28) together lead to,
\be \gamma\frac{h'}{h} = -m,\;\;\;  \xi = -\frac{2m}{n_1}.\ee
Thus, $\xi$ turns out to be a pure constant and hence, $\beta = \xi\beta_{3}(R)$ turns out to be a function of $R$ only. It can be checked that in view of these solutions, condition (25) is trivially satisfied. We are finally left with equation (13), which can be expressed as,
\be m\Big(F - \frac{6k}{a^2}F_{,R}\Big)-\frac{1}{2 h}\left(3 m V  + \gamma V'\right) = 0.\ee
Apparently, equation (30) is satisfied only if $V$ and $h$ are constants, while $k = 0$. But then equations (27) and (28) are satisfied simultaneously, for $m = 0$, which makes $\alpha = 0 = \beta$ and satisfies equation (30) trivially keeping the form of $F(R)$ obscure. Summarily, we find that, the Noether equations (7) through (13) can not be solved for a nonlinear form of $F(R)$ and so we fail to extract Noether symmetry, at least following the technique of treating $R - 6(\frac{\ddot a}{a} + \frac{\dot a^2}{a^2} + \frac{k}{a^2}) = 0$ to be a constraint of the theory. Situation does not improve under the addition of dark and baryonic matter. To be more conclusive, we take up a more general action by adding the Ricci scalar in the following section.

\section{$F(R)$ in addition to non-minimally coupled scalar tensor theory of gravity.}
Here, we take up even a more general form of action, in which $F(R)$ is added to a non-minimally coupled scalar-tensor theory of gravity, as

\be S = \int\sqrt{-g}d^4 x\left[f(\phi)R + B F(R) - \frac{\omega(\phi)}{2}\phi_{,\mu}\phi^{,\mu} - V(\phi)\right].\ee
Note that this is the same action for which Noether symmetry was found earlier taking $F(R) = R^2$ a-priori and following the standard technique of using an auxiliary variable $Q$ different from $R$, such that the Lagrangian was spanned by the configuration space variables $(a, Q, \phi, \dot a, \dot Q, \dot\phi)$ \cite{8}. So, we expect to get Noether symmetry for the above action (31) at least for $F(R) = R^2$, following the technique of treating $R - 6(\frac{\ddot a}{a} + \frac{\dot a^2}{a^2} + \frac{k}{a^2}) = 0$ as a constraint of the theory. This is what we pose in this section. Substituting the expression of the Lagrange multiplier and removing appropriate surface term, the point Lagrangian corresponding to the action (31) in the Robertson-Walker space time (2) reads,

\be L = 6(-f\dot a^2-f'a\dot a\dot \phi+kf)a+B(F-R F_{,R})a^3-6B(F_{,R}\dot a^2+a\dot a\dot R F_{,RR}-k a F_{,R})+a^3\big(\frac{\omega\dot\phi^2}{2}-V\big).\ee

\noindent Noether equation are,

\begin{eqnarray}
&(\alpha+2a\alpha_{,a})(B F_{,R}+f)+B(\beta+a\beta_{,a})a F_{,RR}+f'a(\gamma+a\gamma_{,a}) = 0.\\
& 6Ba^2\alpha_{,R}F_{,RR} = 0.\\
& 3(\omega\alpha-4f'\alpha')+a(\gamma\omega'+2\gamma'\omega) = 0.\\
& B\beta a F_{,RRR} + B(2\alpha+a\alpha_{,a}+a\beta_{,R})F_{,RR}+f'a\gamma_{,R} + 2(f + B F_{,R})\alpha_{,R}=0.\\
& 6BF_{,RR}\alpha' - \omega a\gamma_{,R} + 6f' \alpha_{,R} = 0.\\
& 12[\alpha  f'+(f+BF_{,R})\alpha']+6[\alpha_{,a}f'+\gamma f''+B\beta'F_{,RR}]a-\omega a^2\gamma_{,a}+6a\gamma'f' = 0.\\
& \alpha[6kf+6kBF_{,R}+3Ba^2(F-R F_{,R})-3a^2V]+\beta[B(-a^3R+6ka)F_{,RR}]+\gamma[6kf'a-a^3V'] = 0.
\end{eqnarray}
Following subsection is devoted to find solutions of the above set of equations (33) through (39) under the condition $F_{,RR} \ne 0$.

\subsection{Solutions.}
Equation (34) implies $\alpha \ne \alpha(R)$, i.e., $\alpha = \alpha(a,\phi)$ as before, since we are interested in non-linear form of $F(R)$. Hence equation (37) leads to two distinct cases, which we explore in the following subsubsections (3.1.1) and (3.1.2).\\
\subsubsection{Case-1}
Since, $\alpha_{,R} = 0$, so equation (37) may be integrated immediately to yield,
\be F_{,R} = \frac{wa}{6B\alpha'}\gamma + n, \Longrightarrow \gamma = \frac{6B\alpha'}{wa}(F_{,R} - n),\ee
where, as before, $n$ is a pure constant, since $F_{,R}$ is only a function of $R$. Substituting the above form of $\gamma$ in equation (35), we get,

\[2B(w'\alpha' - 2w\alpha'')(F_{,R}-n) = w^2\alpha - 4f'w\alpha',\hspace{3.54in}(vi).\]
The above equation is satisfied if both sides separately vanish, which is possible under the following conditions,

\[2w\alpha'' =w'\alpha' \Longrightarrow \alpha'^2 = \alpha_1(a)^2 w,\hspace{4.22in}(vii)\]
\[w^2\alpha = 4f'w\alpha'.\hspace{5.32in}(viii)\]

\noindent Equations $(vii)$ and $(viii)$ imply,
\[ \frac{\alpha\alpha'}{\alpha_{1}^2} = 4f'.\hspace{5.59in}(ix)\]

\noindent Hence $\alpha$ must admit the separable form, $\alpha = \alpha_1(a)\alpha_2(\phi)$, and so,
\be
\alpha_2^2 = 8f,\;\;\;\;and\;\;\;w = \alpha_2'^2.
\ee

\noindent Now, using equations $(vii)$ and $(viii)$, in equation (32) following equation is found,
\be
B(\xi + a \xi_{,a})a\beta_3 F_{,RR} + B F_{,R}\left[(\alpha_1+2a \alpha_{1,a})\alpha_2+\frac{6f'a\alpha_{1,a}\alpha_2'}{\omega}\right]+f(\alpha_1+2a \alpha_{1,a})\alpha_2-\frac{6Bnf'a\alpha_{1,a}\alpha_2'}{\omega}= 0.
\ee
\noindent
where, $\beta = \xi(a, \phi)\beta_3(R)$. Equation (42) is satisfied under different conditions which we discuss in the following three subcases.\\

\textbf{Subcase I.}
\begin{eqnarray}
&\frac{\beta_3F_{,RR}}{F_{,R}} = l,\\
& l B (\xi+a\xi_{,a})a + B \left[ (\alpha_1+2a \alpha_{1,a})\alpha_2+\frac{6f'a\alpha_{1,a}\alpha_2'}{\omega}\right]=0,\\
& f(\alpha_1+2a \alpha_{1,a})\alpha_2-\frac{6Bnf'a\alpha_{1,a}\alpha_2'}{\omega} = 0
\end{eqnarray}

\noindent where, $l$ is a constant. Now using equation (41), one can express equation (45) as

\be
\alpha_2^3(\alpha_1+2a \alpha_{1,a})-12 B n  a \alpha_{1,a}\alpha_2=0.
\ee
Here, we consider $\alpha_2'\neq 0$ because $\alpha_2' = 0$ imply $\omega = 0$ as may be seen from equation (41). Thus, equation (46) is satisfied only under the conditions,

\[\alpha_1+2a \alpha_{1,a}=0\;\;\; and\;\;\; \alpha_{1,a}=0.\]

\noindent The above two equations imply $\alpha_1 = 0$, which leads to a trivial case. Hence, we take up the other alternative in the following subcase.\\

\textbf{Subcase II.}
Equation (42) is also satisfied provided,
\begin{eqnarray}
&\beta_3F_{,RR}=l=const,\\
& l B (\xi+a\xi_{,a})a + f(\alpha_1+2a \alpha_{1,a})\alpha_2-\frac{6Bnf'a\alpha_{1,a}\alpha_2'}{\omega} = 0,\\
& (\alpha_1+2a \alpha_{1,a})\alpha_2+\frac{6f'a\alpha_{1,a}\alpha_2'}{\omega}=0
\end{eqnarray}

\noindent Again, using equation (41) we have from equation (49)

\be
2\alpha_1+7a \alpha_{1,a}=0.
\ee
Now, in view of equation (47) one gets, $\beta_3F_{,RRR}+\beta_{3,R}F_{,RR}=0.$ Using this together with equations (40) and (41), equation (36) takes the form,

\be
7\alpha_1+2a\alpha_{1,a}=0.
\ee

\noindent Equations (50) and (51) are simultaneously satisfied only for $\alpha_1 = 0$, which again leads to trivial situation.\\

\textbf{Subcase III.}\\

\noindent
Here, we consider that the coefficients of equation (42) vanish separately, i.e.,
\[(\xi + a \xi_{,a})\beta_3=0,\;\;\;
(\alpha_1+2a \alpha_{1,a})\alpha_2+\frac{6f'a\alpha_{1,a}\alpha_2'}{\omega}=0,\;\;\;
f(\alpha_1+2a \alpha_{1,a})\alpha_2-\frac{6Bnf'a\alpha_{1,a}\alpha_2'}{\omega}=0.\]

\noindent Note that, the last equation is the same as equation (45) of subcase I, which only leads to trivial result. Thus the case under consideration does not yield any symmetry for a nonlinear form of $F(R)$.
\subsubsection{Case 2.}
Equation (37) is also satisfied under the condition, {$\alpha' = 0 =\gamma_{,R}$}, i.e., $\alpha=\alpha(a)$ and $\gamma=\gamma(a,\phi)$. Therefore, equation (33) may be expressed as,
\[\frac{3\alpha}{a\gamma_1}+\gamma_2\frac{\omega'}{\omega}+2\gamma_2'=0,\] where, we have separated  $\gamma(a,\phi)=\gamma_1(a)\gamma_2(\phi)$.
\noindent The above equation is true only if,
\be
\frac{3\alpha}{a\gamma_1}=c_1~~and~~\gamma_2\frac{\omega'}{\omega}+2\gamma_2'=-c_1
\ee
\noindent $c_1$ being a constant.
\noindent Now, separating $\beta$ as, $\beta=\xi(a,\phi)\beta_3(R)$, equation (31) leads to
\be
B(\xi + a \xi_{,a})a\beta_3 F_{,RR} + B F_{,R}(\alpha+2a \alpha_{,a})+f(\alpha+2a \alpha_{,a})+f'a(\gamma+a\gamma_{,a})= 0.
\ee
\noindent Note that, in equation (53) the condition $\beta_3 F_{,RR} = $ constant, automatically implies $F_{,R} = $ constant, which is not our concern, so equation (53) is satisfied under the following conditions, viz., \\

\textbf{Subcase I.}\\
\begin{eqnarray}
&\beta_3F_{,RR}=c F_{,R},\\
& c (\xi+a\xi_{,a})a + (\alpha+2a \alpha_{,a}) = 0,\\
& f(\alpha+2a \alpha_{,a})+f'a(\gamma+a\gamma_{,a})= 0,
\end{eqnarray}
c being a constant. Now, using equation (54), equation (39) can be rewritten as
\be -B a^2(c a \xi+3\alpha)RF_{,R}+6kB(c a\xi+\alpha)F_{,R}+3Ba^2\alpha F+\alpha(6kf-3a^2V)+\gamma_1\gamma_2(6kf'a-a^3V')=0\ee
\noindent Here, none of $\alpha$, $\xi$, $\gamma_1$ and $\gamma_2$ are function of $R$ and so, the above equation is satisfied provided the coefficients of different derivatives of $F$ vanish separately. This yields a trivial result, viz., $\alpha = \beta = \gamma = 0$. Hence we turn our attention to the following subcase.\\

\textbf{Subcase II.}\\
Equation (53) is also satisfied under the following set of conditions, viz.,
\begin{eqnarray}
&\beta_3F_{,RR}=c,\\
& c B (\xi+a\xi_{,a})a + f(\alpha+2a \alpha_{,a})+f'a(\gamma+a\gamma_{,a})=0,\\
&\alpha+2a \alpha_{,a}= 0.
\end{eqnarray}
\noindent But then, using equation (58) along with the condition $\gamma_{,R} = 0$, equation (36) is expressed as,
\be
2\alpha+a\alpha_{,a}=0
\ee
\noindent equations (60) and (61) together imply that $\alpha=0$, which leads to trivial situation as discussed earlier. Thus we fail to extract a non-linear form of $F(R)$, imposing Noether symmetry in the most general form of an action. Here again the situation remains unaltered under the addition of dark and baryonic matter. It appears, that the technique of using the form of the Ricci scalar $R - 6(\frac{\ddot a}{a} + \frac{\dot a^2}{a^2} + \frac{k}{a^2}) = 0$ as a constraint in the action is wrong, since as repeatedly mentioned, noether symmetry for such a theory with $F(R) = R^2$ already exists in the literature \cite{18}.

\section{In search of Noether symmetry of $F(R)$ in anisotropic models.}

In this section we proceed to apply the same technique in some of the anisotropic models, taking into account pure $F(R)$ theory of gravity. As in the case of Robertson-Walker metric, we can expect some form of $F(R)$ here too. The form may be different, since the constraint is different. For, this purpose, we take up spatially symmetric Kantowski-Sachs (K-S), Bianchi-I (B-I) and Bianchi-III (B-III) metric, which can be expressed altogether as,

\be ds^2 = -dt^2 + a^2 dr^2 + b^2[d\theta^2 + f_k^2 d\phi^2], \ee

where, $f_k = sin\theta \Rightarrow k = + 1$ (K-S), $f_k = \theta \Rightarrow k = 0$ (B - I), $f_k = sinh\theta \Rightarrow k = - 1$ (B - III), and for which

\be ^4R = 2\left(\frac{\ddot a}{a}+2\frac{\ddot b}{b}+2\frac{\dot a\dot b}{ab}+\frac{\dot b^2}{b^2} +\frac{k}{b^2} \right).\ee

\noindent
Under variation of the action

\be A = \int\left[B F(R) - \lambda\left\{^4R - 2\left(\frac{\ddot a}{a}+2\frac{\ddot b}{b}+2\frac{\dot a\dot b}{ab}+\frac{\dot b^2}{b^2} +\frac{k}{b^2} \right)\right\}\right]\sqrt{-g} dt,\ee

\noindent
with respect to $R$ yields, the same old form, viz., $\lambda = BF_{,R}$. Thus the action after removing total derivative terms yields,

\be A = \int\left[ab^2(F - R F_{,R})-2F_{,R}(2b\dot a\dot b + a\dot b^2 - k a) - 2 F_{,RR}(b^2\dot a + 2ab\dot b)\dot R\right]dt + \Sigma,\ee

\noindent
where, $\Sigma = 2F_{,R}(\dot ab + 2a\dot b) b,$ which is at par with the surface term obtained from the standard variational technique. Now, in order to find a form of $F(R)$, if we impose Noether symmetry ($\pounds_X L = X L = 0$) , we get following equation, viz.,

\[ \alpha[b^2(F - R F_{,R})-2F_{,R}(\dot b^2-k)-4F_{,RR}b\dot b\dot R]+\beta[2ab(F-RF_{,R})-4F_{,R}\dot a\dot b-4F_{,RR}(b\dot a+a\dot b)\dot R]+\]
\[\gamma\left[-F_{,RR}[Rab^2+2(2b\dot a\dot b+a\dot b^2-ka)]-2\dot R\dot{\overbrace{(ab^2)}}F_{,RRR}\right]- \left(\frac{\partial\alpha}{\partial a}\dot a+\frac{\partial\alpha}{\partial b}\dot b+\frac{\partial\alpha}{\partial R}\dot R\right)[4F_{,R}b\dot b+2F_{,RR}b^2\dot R]-\]
\be\left(\frac{\partial\beta}{\partial a}\dot a+\frac{\partial\beta}{\partial b}\dot b+\frac{\partial\beta}{\partial R}\dot R \right)[4F_{,R}(b\dot a+a\dot b)+4F_{,RR}ab\dot R)]-\left(\frac{\partial\gamma}{\partial a}\dot a+\frac{\partial\gamma}{\partial b}\dot b+\frac{\partial\gamma}{\partial R}\dot R \right)[2F_{,RR}(b^2\dot a+2ab\dot b)]=0.\ee
So, as usual, equating coefficients of $\dot a^2, \dot b^2, \dot a\dot b, \dot R^2, \dot b\dot R, \dot R\dot a$ and others to zero, following seven Noether  equations are found, viz.,
\begin{eqnarray}
&\frac{F_{,RR}}{F_{,R}}=-\frac{2\beta_{,a}}{b\gamma_{,a}}. \\
&\frac{F_{,RR}}{F_{,R}}=-\frac{\alpha+2b\alpha_{,b}+2a\beta_{,b}}{a(\gamma+2b\gamma_{,b})}. \\
&\frac{F_{,RR}}{F_{,R}}=-2\frac{b\alpha_{,a}+\beta+a\beta_{,a}+b\beta_{,b}}{b(2\gamma+2a\gamma_{,a}+b\gamma_{,b})}. \\
&b\alpha_{,R} + 2a \beta_{,R}=0. \\
&\frac{F_{,RRR}}{F_{,R}}+\frac{F_{,RR}}{F_{,R}}\left[\frac{2b\alpha+b^2\alpha_{,b}+2a\beta+2ab\beta_{,b}
   +2ab\gamma_{,R}}{2ab\gamma}\right]+\frac{b\alpha_{,R}+a\beta_{,R}}{ab\gamma}=0. \\
&\frac{F_{,RRR}}{F_{,R}}+\frac{F_{,RR}}{F_{,R}}\left[\frac{2\beta+b\alpha_{,a}+2a\beta_{,a}+b\gamma_{,R}}{b\gamma}\right]
   +2\frac{\beta_{,R}}{b\gamma}=0. \\
&F_{,RR}+\left[\frac{b^2R\alpha-2k\alpha+2abR\beta}{a(b^2 R - 2k)\gamma}\right]F_{,R}-\left[\frac{b(b\alpha+2a\beta)}{a(b^2 R -2k)\gamma}\right]F =0.
\end{eqnarray}

\subsection{Solution of Noether equations.}

In order to find a non-linear form of $F(R)$, we are required to solve the above set of Noether equations containing four variables $\alpha, \beta, \gamma$ and $F$, under the assumption $F_{,RR} \ne 0$. For this purpose, let us separate the variables as,

\be \alpha = A(a,b)D_1(R),\;\;\;\; \beta = B(a,b)D_2(R), \;\;\;\;\gamma = C(a,b)D_3(R).\ee
In view of the above choices made in (74), equations (67), (68), (69), (70) and (73) take the following form respectively,

\begin{eqnarray}
& \frac{D_3F_{,RR}}{D_2F_{,R}} = n_1 = -2\frac{B_{,a}}{b C_{,a}},\\
& \frac{D_3F_{,RR}}{F_{,R}} = -\left[\frac{(A+2bA_{,b})D_1 + 2aB_{,b}D_2}{a(C + 2bC_{,b})}\right],\\
& \frac{D_3F_{,RR}}{F_{,R}} = -2\left[\frac{b A_{,a}D_1 + (B + a B_{,a} + b B_{,b})D_2}{b (2C + 2a C_{,a} + b C_{,b})}\right],\\
& b A D_{1,R} + 2 a B D_{2,R} = 0.\\
& D_3 F_{,RR} + \left[\frac{b(b A D_1 + 2 a B D_2)(R F_{,R} - F) - 2kA D_1 F_{,R}}{a C(b^2 R - 2k)}\right] = 0,
\end{eqnarray}
where, $n_1$ is a separation constant. In view of (76) and (77), one immediately gets

\be D_2 = m_1 D_1,\ee
In view of which equation (78) now reads,

\be  b A + 2 m_1 a B = 0,\ee
where, $m_1$ is another separation constant. Replacing $F_{,RR}$ by $F_{,R}$ and using equations (80) and (81) in equation (79) one gets,

\be \left[m_1 n_1 - \frac{2k A}{a(b^2 R - 2k)C}\right]F_{,R} = 0.\ee
Since $F_{,R} \ne 0$, so the above equation clearly yields a solution of the curvature scalar as,

\be R = \frac{2k}{b^2}\left[\frac{A}{m_1 n_1 a C} + 1\right] = n_2,\ee
where, $n_2$ is yet another separation constant. Thus $R$ turns out to be a constant, which is zero for $k = 0$. This result directly contradicts higher order theory of gravity. The field equation for higher order theory of gravity is,

\be F,_RR_{\mu\nu}-\frac{1}{2}F g_{\mu\nu}+ {{(F,_R)}^{;\alpha}}_{;\alpha}~ g_{\mu\nu}-(F,_R)_{;\mu;\nu} =0 .\ee
Trace of which yields,

\be R F,_R+3{{(F,_R)}^{;\alpha}}_{;\alpha}-2F = 0.\ee
Thus, $R$ can not be zero or constant, which implies Birkhoff's theorem that Schwarzschild's solution is the unique spherically symmetric vacuum solution does not hold for $F(R)$ theory of gravity. In fact, substituting the above result that $R$ is a constant in equation (79), one obtains

\be F_{,RR} = 0,\ee
which contradicts the starting assumption that $F_{,RR} \ne 0$. Thus, we observe that Noether symmetry of $F(R)$ theory of gravity fails to yield a non-linear form of $F(R)$ in the anisotropic models under consideration. Thus, despite rigorous calculation, Noether symmetry of $F(R)$ theory of gravity remains obscure.
This is not a problem whatsoever, since nature may not allow such additional symmetry. Problem is, Noether symmetry for $F(R) = R^2$ already exists \cite{18} and the present technique \cite{21} fails to reproduce it. Hence, fixing $F(R)$ a-priori and expressing the canonical form of the Lagrangian using auxiliary variable $Q$ \cite{22a}, Noether symmetry may be realized. On the other hand, since there is no way to find a canonical point Lagrangian for $F(R)$ theory of gravity, other than the technique under consideration \cite{21}, hence at this end we conclude that Noether symmetry of $F(R)$ theory of gravity is obscure and so there is no way to find a form of $F(R)$, other than choosing it arbitrarily.

\section{A general conserved current for $F(R)$ theory of gravity.}

 Noether symmetry not only fixes the form of a parameter but also yields an integral of motion leading to a cyclic variable and thus helps in finding some exact solutions of apparently non-tractable field equations. Thus absence of Noether symmetry for $F(R)$ gravity makes it practically impossible to solve the field equations even if a form of $F(R)$ is put in by hand. In this section we show that the situation is not as bad as it appears, because a general (Non-Noether) conserved current exists for higher order theory of gravity \cite{27}. For the purpose, here we take up action (31) in addition to a matter Lagrangian $L_{m}$, to briefly illustrate the existence of the said integral of motion, as

\be
A = \int\left[f(\phi) ~R + B F(R)
-\omega(\phi)~\phi,_{\mu}\phi^{,\mu} -V(\phi)-\kappa
L_{m}\right]\sqrt{-g}~d^4x. \ee

\noindent
Field equations corresponding to the above action under metric variation method are,

\[
f\left(R_{\mu\nu}-\frac{1}{2}g_{\mu\nu}R\right)+{f^{;\alpha}}_{;\alpha}~g_{\mu\nu}-f_{;\mu
;\nu} -\omega~\phi_{,\mu}\phi_{,\nu}+
\frac{1}{2}g_{\mu\nu}\Big(\omega\phi_{,\alpha}\phi^{,\alpha}+V(\phi)\Big)\]
\be +B\left[(F,_R)R_{\mu\nu}-\frac{1}{2}F g_{\mu\nu}+
{{(F,_R)}^{;\alpha}}_{;\alpha}~ g_{\mu\nu}-(F,_R)_{;\mu
;\nu}\right]=\frac{\kappa}{2}T_{\mu\nu}. \ee

\be
R f'+2\omega\phi^{;\mu}_{;\mu}+\omega'\phi^{,\mu}\phi_{,\mu}-V'(\phi) = 0.
\ee
The trace of the field equation (88) is,
\be R f-3f^{;\mu}_{;\mu}-\frac{\omega}{\phi}\phi^{,\mu}\phi_{,\mu}-2V -B[R(F,_R)+3{{(F,_R)}^{;\alpha}}_{;\alpha}-2F] = -\frac{\kappa}{2}T^{\mu}_{\mu}. \ee
Combining equations (89) and (90) it has been shown \cite{18} that under the following condition,

\be B [R(F,_R) +3{{(F,_R)}^{;\alpha}}_{;\alpha}-2F] =
\frac{\kappa}{2}~T^{\mu}_{\mu} + \frac{f^3}{f'}\left(\frac{V}{f^2}\right)', \ee
the above action (87) carries a conserved current

\be
J^{\mu}_{;\mu}=[(3f'^2+2f \omega)^{\frac{1}{2}}\phi^{;\mu}]_{;\mu}
= 0, \ee

\noindent
which in the case of homogeneous cosmology, takes the following simplified form,

\be \sqrt{-g\left(3f'^2+2f~\omega\right)}\;\;\dot \phi = c, \ee

\noindent
where, $c$ is a non-vanishing constant and $g$ is the determinant of the metric. Thus, a general conserved current exists for higher order theory of gravity, without even fixing the forms of the parameters involved, which lies beyond the scope of Noether symmetry, since it is true for the theory of gravitation in general.
\subsection{Attempting a solution for $F(R) = R^{\frac{3}{2}}$.}
It might still appear impossible to solve the cosmological field equations using the conserved current (93). However, it has been expatiated in a recent letter \cite{28}, that it may be useful to solve the field equations taking $F(R) = R^2$. Here, just to cite another example we try the same for $F(R) = R^{\frac{3}{2}}$. Remember that $F(R) = R^{\frac{3}{2}}$ yields a Noether symmetry in the Robertson-Walker space-time only in the absence of a linear term and the dilatonic sector. However, here we are presenting a solution corresponding to an action containing a linear term being non-minimally coupled to a scalar field in the presence of $F(R) = R^{\frac{3}{2}}$. In the Robertson-Walker space time, we are required to solve the four available equations, for $a(t), \phi(t), f(\phi), \omega(\phi)$ and $V(\phi)$. Thus one has to make yet another assumption. To handle the equations comfortably, we therefore split the condition (91) into two, assuming $V \propto f(\phi)^2$, as

\be B [R F,_R +3{{(F,_R)}^{;\alpha}}_{;\alpha}-2F] = \frac{\kappa}{2}~T^{\mu}_{\mu},  \ee

\be V = \lambda f^2,\ee

\noindent
where, $\lambda$ is a constant. So finally, in the case of homogeneous cosmology, we need to solve following set of five equations,
\begin{eqnarray}
& B [R F,_R +3\Box F,_R-2F] = \frac{\kappa}{2}~T^{\mu}_{\mu},  \\
& V = \lambda f^2,\\
& R f-3 \Box f -\omega\phi^{,\mu}\phi_{,\mu}-2V=0, \\
& R f'+2\omega\Box\phi+\omega'\phi^{,\mu}\phi_{,\mu}-V'(\phi) = 0,\\
& \sqrt{-g\left(3f'^2+2f~\omega\right)}\;\;\dot \phi = c.
\end{eqnarray}

\noindent  Now for $F(R) = R^{\frac{3}{2}}$,
\be \Box F_{,R} = -[(\ddot R + \Theta \dot R)F_{,RR} + \dot R^2 F_{,RRR}] =  -\left[\frac{3}{4}\frac{\ddot R}{\sqrt R} -\frac{3}{8}\frac{\dot R^2}{R^{\frac{3}{2}}} + \frac{27}{4}H\frac{\dot R}{\sqrt R} ,\right],\ee

\noindent
and so equation (96) reduces to,

\be \left[9\frac{\ddot R}{\sqrt R} -\frac{9}{2}\frac{\dot R^2}{R^{\frac{3}{2}}} + 9H\frac{\dot R}{\sqrt R}+ 2 R^{\frac{3}{2}}\right] = -\frac{2\kappa}{B}~T^{\mu}_{\mu}.\ee

\noindent
Taking Baryonic and non-Baryonic dark matter in the energy momentum tensor, one can solve the above equation (102), both in the radiation and matter dominated era in the flat case $k = 0$. In the radiation dominated era, taking $p = \frac{1}{3}\rho$, the trace of the energy momentum tensor $T_{\mu}^{\mu}$ vanishes and equation (102), gets solved to yield $a \propto \sqrt t$. This is a brilliant feature of higher order theory of gravity, that despite being tightly coupled to the scalar field it still produces Friedmann type solution in the radiation era keeping standard big-bang-nucleosynthesis (BBN) unchanged. Note that we are required to find $\phi, f(\phi), V(\phi) $ and $\omega({\phi})$, in view of the four equations (97) through (100), which although we are not attempting, may be found in principle, since explicit form of $R$ is now known. In the matter dominated era, on the other end, $p = 0$ and the trace of the energy momentum tensor is simply $T_{\mu}^{\mu} = \rho$, with $a^3 \rho = \rho_{m0}$, $\rho_{m0}$ being the present matter density of the Universe. The only solution that we could find for equation (102) is $a = a_0 t$, provided the condition $B = -\frac{\sqrt6\kappa\rho_{m0}}{90 a_0^3}$ holds. However, such a coasting solution although fits SnIa data perfectly as illustrated in figure-I, is ruled out, since it does not admit structure formation. In our earlier work \cite{22}, we found that $R^{\frac{3}{2}}$ yields $a \propto t^{\frac{3}{4}}$ in the radiation era, which creates problem in explaining BBN and its solution in the matter dominated era, although admits early deceleration and late time acceleration, does not fit with the Hubble parameter versus red-shift - standard model data. Therefore, we have tried here to incorporate a linear term in the action, being coupled with a dilatonic sector. It shows, despite improvement in radiation era, it does not admit experimental data in the matter dominated era. Hence, $R^{\frac{3}{2}}$ term is likely to be ruled out.

\begin{figure}
[ptb]
\begin{center}
\includegraphics[
height=2.034in, width=2.8in] {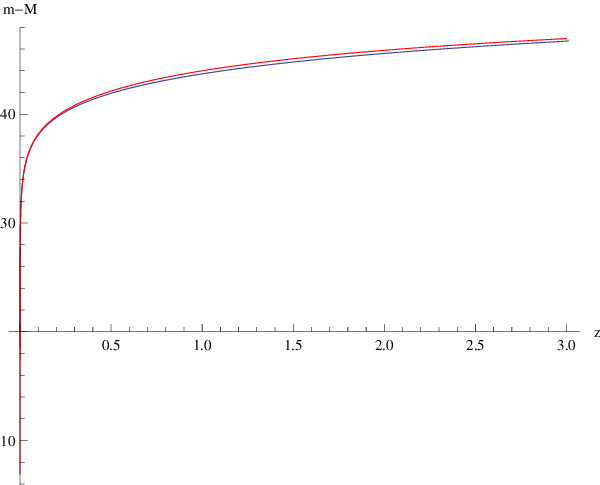} \caption{Distance modulus $(M-m)$ versus redshift $z$ plot of the present model (blue) satisfying coasting solution $a \propto t$, shows perfect fit with the $\Lambda$CDM model (red).}
\end{center}
\end{figure}

\section{Summary.}

Both the metric and Palatini variation methods have vices and virtues and it is a matter of taste to adopt one. However at the end one requires a form of $F(R)$, any nonlinear form of which makes field equations extremely difficult to solve. Noether symmetry not only fixes a form out of indefinitely many, but also yields an integral of motion which makes field equations somewhat tractable. For such purpose one requires a canonical point Lagrangian. The only technique known so far to express $F(R)$ in such a form is to introduce $R - 6(\frac{\ddot a}{a} + \frac{\dot a^2}{a^2} + \frac{k}{a^2}) = 0$ as a constraint in Robertson-Walker metric, which gives correct field equations, in view of the auxiliary variable $R$. Following this technique, Noether symmetry yields $F(R) = R^{\frac{3}{2}}$ both in vacuum and matter dominated era. Although such a form is reasonably good in the early Universe, suffers from the disease that it does not produce standard Friedmann radiation era \cite{22}. In anticipation that a linear term in addition could have cured the disease, here we have enlarged the configuration space by adding a scalar field. Rigorous and detailed calculation shows that Noether symmetry remains obscure for $F_{,RR} \ne 0$. We have also attempted to find Noether symmetry for pure $F(R)$ gravity in the background of some anisotropic models and ended up with the same result. Nature might not allow additional symmetry, but the problem is that Noether symmetry for a scalar-tensor theory of gravity being coupled with $F(R)  = R^2$, already exists in the literature \cite{18}, while the present technique fails to reproduce the same. This raises doubt about the technique \cite{21} under consideration.\\

\noindent
Obviously, one should ask ``how come the technique then yields $F(R) = R^{\frac{3}{2}}$ in vacuum and matter dominated era?". The answer has been given in a recent work \cite{22}, where it has been shown that under appropriate choice of co-ordinate $h_{ij} = a^2 =z$, an action containing $R^{\frac{3}{2}}$ in vacuum and in the matter dominated era, makes the variable $z$ cyclic. Thus, the symmetry obtained in Robertson-Walker space-time yielding $F(R) = R^{\frac{3}{2}}$, is due to the very speciality of the isotropic and homogeneous space-time and so is a mere accident. Since there is no way to find a canonical point Lagrangian for $F(R)$ theory of gravity, other than the one under consideration, so we conclude that such a symmetry is obscure for a general $F(R)$ gravity.\\

\noindent
It therefore appears that there is practically no way to handle $F(R)$ theory of gravity. However, the situation is not as bad, since a scalar-tensor theory of gravity in the presence of $F(R)$, indeed carries a general conserved current \cite{27,28}, which is independent of the metric and also on the forms of parameters involved, viz., $F(R)$, $f(\phi)$, $\omega(\phi)$ and $V(\phi)$. Here in section (5), we have briefly presented the conserved current and cited an example taking $F(R) = R^{\frac{3}{2}}$, to understand its use in finding solutions for higher order curvature invariant terms. The presence of a linear term in the action modifies the solution in the radiation era to the standard Friedmann type ($a \propto \sqrt t$). Thus BBN remains unaltered, which is of-course a brilliant result. On the other hand matter dominated era admits only a coasting solution in the form $a \propto t$, which although fits SNIa data perfectly, does not allow a transition from early deceleration and so $R^{\frac{3}{2}}$ is not a good choice to explain late time cosmological observations.  \\

\end{document}